\documentclass[journal]{IEEEtran}

\ifCLASSINFOpdf
\else
\fi

\usepackage{amsmath}
\usepackage{amsfonts}

\usepackage{nomencl}
\makenomenclature

\usepackage[ruled,vlined]{algorithm2e}

\usepackage[table]{xcolor}
\usepackage{makecell}
\usepackage{multirow}

\usepackage{graphicx}
\usepackage{caption}
\usepackage{subcaption}
\usepackage{svg}
\usepackage{framed}

\usepackage[
backend=biber,
style=ieee,
sorting=none
]{biblatex}
\usepackage{hyperref}

\begin{document}

\title{pmuBAGE: The Benchmarking Assortment of Generated PMU Data for Power System Events\\
}

\author{Brandon~Foggo,~\IEEEmembership{Member,~IEEE,}
        Koji Yamashita,~\IEEEmembership{Member,~IEEE,}
        Nanpeng~Yu,~\IEEEmembership{Senior Member,~IEEE,}
}

\newcommand{\dee}[2]{\frac{\partial{#1}}{\partial{#2}}}

\maketitle

\begin{abstract}

This paper introduces pmuGE (phasor measurement unit Generator of Events), one of the first data-driven generative model for power system event data. We have trained this model on thousands of actual events and created a dataset denoted pmuBAGE (the Benchmarking Assortment of Generated PMU Events). The dataset consists of almost 1000 instances of labeled event data to encourage benchmark evaluations on phasor measurement unit (PMU) data analytics. 
PMU data are challenging to obtain, especially those covering event periods. Nevertheless, power system problems have recently seen phenomenal advancements via data-driven machine learning solutions. 
A highly accessible standard benchmarking dataset would enable a drastic acceleration of the development of successful machine learning techniques in this field. We propose a novel learning method based on the Event Participation Decomposition of Power System Events, which makes it possible to learn a generative model of PMU data during system anomalies. The model can create highly realistic event data without compromising the differential privacy of the PMUs used to train it. The dataset is available online for any researcher or practitioner to use at the pmuBAGE \href{https://github.com/NanpengYu/pmuBAGE}{Github Repository}.

\end{abstract}


%
\IEEEpeerreviewmaketitle

~

\section{Introduction}
 
\IEEEPARstart{S}{olutions} to any power system dynamic study require realistic dynamic response data. 
Obtaining realistic phasor measurement unit (PMU) data that are used by the transmission system operator (TSO) 
or electric utilities has always been a bottleneck in academia. 
Large-scale IEEE dynamic test cases, such as the 145-bus system \cite{IEEE50}, have been widely leveraged to cope with this. 
However, generally accepted dynamic models, such as combined cycle gas turbine models, and renewable energy sources (RES), are missing in such dynamic test cases. 
Although generic dynamic models for turbine-governors \cite{PES_TR1} have been publicly available and implemented in many commercial dynamic simulation software, the widely accepted model parameters are not fully provided.

The power system engineering research center has begun to create synthetic dynamical data by simulating more realistic North American Eastern and Western Interconnection models \cite{PSERC2021}. 
Many non-data driven modelling approaches to create synthetic data can be found in literature. Much of this research builds on each other in a modular way. For example, reference \cite{7515182} uses geographic properties to automatically generate synthetic networks
. Reference \cite{8334287} automatically adjusts the parameters of the generators according to fuel type and the statistics of real generators in the same geographic region. Reference 
\cite{9174809} 
focuses on improving load properties in synthesized grids.
Reference \cite{9216129} focuses on integrating synthetic transmission systems with synthetic distribution systems.
Finally, reference \cite{idehen2020large} performs dynamic simulations on a synthetic Texas power grid to generated synthetic PMU data.

The results of these attempts surely help demonstrate some particular dynamic aspects of power system events more accurately - such as poorly damped 
oscillations and frequency drop following disturbances. 
However, we have no all-encompassing or one-size-fits-all dynamic model. 
Besides, the parameterization of the aforementioned generic model, specifically RES models, becomes more difficult due to its accelerating spatiotemporal dispersion, which results in insufficient simulation accuracy (i.e., less credible simulation results), especially for the future grid analysis with high penetration of RES. 
Superimposing realistic noise to the simulated response is also not trivial \cite{Kaveri2021Noise}. 
Therefore, real-world measurement data is craved by both academics and industry engineers.

However, access to real-world PMU data is quite limited. Although some TSOs have started to provide sample PMU data to researcher, most do not offer to serve them. The primary factor of hesitation for the release of this data is the potential risk of exposing any vulnerabilities in the bulk power system. Most of the time, power system researchers will go through a lengthy process to setup collaborations with the utility companies who own this PMU data. However, the amount of data shared with the researchers is often small and spanning a somewhat random time-span. There are two meta-level effects to this process. The first effect is that initiating a project on actual data in this field has a considerable time overhead, limiting the amount of quality research that comes out. The second effect is significant heterogeneity of results on the same problem. 


Realistic generated data, on the other hand, is more readily accepted for release by TSOs because such data is no longer treated as real. By generated data, we refer specifically to the process of creating synthetic data by sending random noise through a deep neural network trained on real data. Such deep generative models promise to boost power system dynamic studies, specifically for event detection and classification with big data analysis. The dataset presented in this paper, pmuBAGE, will bring back the advantages of using standard IEEE dynamic test cases - i.e., the accessibility and homogeneity of results - while maintaining the realism and difficulty of dealing with real PMU data.

pmuBAGE is the result of training a novel generative model, called the pmuGE (the phasor measurement unit Generator of Events), and training it on over $1000$ real events gathered over two years on the bulk U.S. Power Grid. While pmuBAGE is an immediately available dataset, the generative model will be described in great detail in this paper. If researchers desire to create an updated version of pmuBAGE using new PMU data, the instructions for doing so are readily available in this paper. The model preserves the privacy of the PMUs used to train it. 

There are a few existing papers on this topic. A simple Generative Adversarial Network (GAN) architecture was trained on PMU data simulated via 9-bus and 39-bus IEEE dynamic test cases \cite{8784681, 9361704}. These two works are important in showing the feasibility of training generative models on PMU data during events. However, the limitations of the data used to train these models - that the corresponding training datasets do not come from real synchrophasor data - resulted in a synthetic datasets with unrealistic noise and PMU-specific behaviors. Example applications of pmuBAGE include experimenting/benchmarking for power system event detection \cite{ brahma2016real, liu2019data, wang2020frequency, zhou2018ensemble, cui2018novel}, event classification \cite{hannon2019real,li2019unsupervised}, and missing value replacement (especially during events) \cite{chatterjee2019robust, liao2018estimate,  osipov2020pmu, konstantinopoulos2020synchrophasor,  huang2016data, 9468345}

To our knowledge, the present work is the first to attempt a fully data driven approach to generating synthetic synchrophasor data using large-scale real world PMU data.

The contributions of this paper are as follows:
\begin{itemize}
    \item By decomposing the PMU data into a static statistical component and a dynamic physical component, separating dynamic components into inter-event and intra-event components, and using 
    probabilistic programming methods and deep cascaded convolutional generative models, the proposed model is able to create much more realistic synthetic event data than was previously possible.
    \item A new QR-reorthogonalization trick is proposed to reintegrate the inter-event signatures with intra event signatures and sparse signatures, which are not typically statistically independent. This significantly increases the ease of training the proposed model.
    \item By introducing several feature matching training loss functions including a completely new loss function denoted as the ``quantile loss", the proposed model is able to capture the distribution of the aforementioned statistical component extremely tightly.
    \item The proposed model is trained on the largest real-world synchrophasor dataset to date.
    This enables the proposed model to capture the PMU-varying and time-varying dynamics of power system event data more accurately.
    \item The first comprehensive set of realistic synthetic events is provided covering many types of event causes, including previously under-modelled event types such as those involving lightning strikes and renewable generation.
\end{itemize}

The remainder of the paper is organized as follows. Section II presents the overall framework of the pmuGE model.  Section \ref{sec:Notation} goes over some notational notes
. Section \ref{sec:EP} details the event-participation decomposition and its computation. Section \ref{sec:Generator} provides all of the information necessary to replicate the generative model for participation factors, with subsections \ref{subsec:overview} through \ref{subsec:losses} detailing that of the intra-event signature generator, and subsection \ref{subsec:inter} detailing those of the inter-event signatures. Section VI shows the numerical study results. Section VII concludes the paper.


 
 


 \begin{figure}
     \centering
     \includegraphics[width=\linewidth]{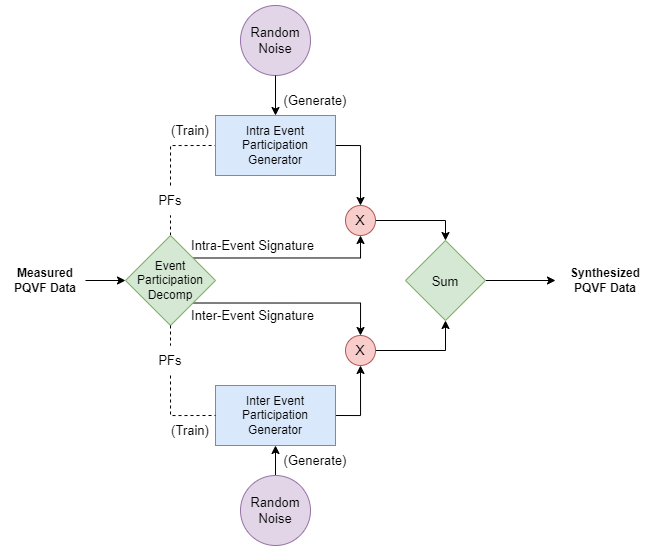}
     \caption{The overall framework of the pmuGE model.}
     \label{fig:overveiw}
 \end{figure}
 \vspace{-0.3cm}
\section{Overall Framework}
 
The goal of this research is to create a realistic PMU dataset for power system events without compromising the privacy of the PMUs used to train it. The proposed ``Event-Participation" (EP) decomposition transformation is the primary device for achieving this. This decomposition separates event tensors into event signatures, shared across all PMUs, and participation factors specific to each PMU \cite{9468345}. By using the Event Participation Decomposition, the data properties that could compromise PMUs are extracted into the participation factors, while the parts that cannot are extracted into event signatures. This allows us to maintain the properties of the dataset that do not vary with PMUs (i.e., the event signatures) in their exact form. Then, a statistical/generative modeling can be performed on the participation factors of this decomposition.
 
A very high-level view of the proposed framework is provided in Figure \ref{fig:overveiw}. Standardized PQVF (real power, reactive power, voltage magnitude, frequency) data tensors are first fed into a module denoted ``EP" (short for Event-Participation Decomposition). The EP module decomposes each tensor into four building blocks - inter-event signatures, intra-event signatures, inter-event participation factors, and intra-event participation factors. No generative modeling is required for either of the event-signature components. On the other hand, the participation factor components cannot be used directly since they carry all of the PMU specific information.

The synthetic Intra-Event components are created via a deep generative probabilistic program resembling a Generative Adversarial Network (GAN). The inter-event participation factors, being much less nuanced in their distributions, are modelled via a statistical simulation.
 
\section{Important Notational Notes}\label{sec:Notation}
 Most of the tensors that appear in this text will have three axes. The first axis will typically denote the datatype - indexing, in order, ``real power," ``reactive power," ``voltage magnitude," and ``frequency." The meaning of the latter two axes will differ by context. However, the products of these tensors will be written with the standard matrix multiplication notation.
 If $A$ and $B$ are both tensors with three axes, then $AB$ is a new three-axis tensor given by contracting the third axis of $A$ with the second axis of $B$. Put another way, $AB$ is the tensor obtained by looping through and holding each index of the first axis of $A$ and $B$ (resulting in $4$ standard matrices each), performing matrix multiplication on each of the holds, and then re-stacking the resulting matrices back into a 3-tensor. For further reference, this is the convention used in the ``matmul" function in both NumPy 
 and PyTorch.
 
\section{The Event-Participation Decomposition}\label{sec:EP}

\subsection{Decomposition Properties and Transformation Tricks}


In this subsection, we give a set of desired properties that the event-participation decomposition ought to have, and present some general tips to obtain such properties. A full description of the decomposition steps will be provided in the next subsection.

\subsubsection{Algorithmic Requirements} There are a myriad of tensor decomposition methods available for use. Selecting which to use depended on two key design constraints. First, the method must have guaranteed convergence over a widely varying set of tensors. Second, the method should yield independent participation factors.

The reason for this first constraint is that the dataset used for training the model involves almost $1000$ power system event tensors, with event causes ranging from downed lines, generator tripping, lightning strikes, and more. A wide variety of dynamic phenomena can be observed in this dataset. Thus, uniform convergence of the chosen tensor decomposition technique is critical. 

The second constraint is not as straightforward as the first. Recall that each participation factor is a set of samples from an event-dependent distribution. Orthogonality means that the distributions corresponding to each co-occurring event signature are statistically uncorrelated. In the ``PMUs-as-samples" viewpoint, this is the only variable observed in multiple instances. Unlike participation factors, event signatures are viewed holistically, not as samples of a random variable. Thus, the independence of event-participation pairs from one another can only be tested by the independence of participation factors.

Independence of event-participation pairs has three important outcomes:
\begin{enumerate}
    \item The distributions are less taxed by the curse of dimensionality, and are therefore easier to learn.
    \item An outlying or otherwise unrealistic sample of one participation factor has no effect on the realism of the other participation factors.
    \item A given power system event signature can be perturbed independently without sacrificing the realism of the others.
\end{enumerate}

The first two of these outcomes depend only on the independence of participation factors. The last outcome relies on assuming that this independence also carries over to the independence of event signatures.

These latter outcomes are expanded upon next. For the second outcome, suppose there are several coinciding event signatures $e_1, e_2, \cdots, e_r$. These event signatures are sent through the learned functions $\mu_{\theta}$ to obtain $\mu_{\theta}(e_1, e_2, \cdots, e_r)=\mu_{\theta_1}(e_1)\mu_{\theta_2}(e_2)\cdots\mu_{\theta_r}(e_r)$. Then $N$ \nomenclature{$N$} values are sampled
from this joint distribution and stacked in vectors to obtain participation factors $p_1, p_2, \cdots, p_r$. \nomenclature{$p$}{Participation Factor} If there is an immense outlying value in $p_1$, then only the outlier in $p_1$ needs to be resampled.
If they are not independent, then all $r$ participation factors need to be resampled.
As the number of co-occurring event signatures increases, the likelihood of one of these participation factors being an outlier increases as well. 

For the third outcome, suppose there are $r$ coinciding event signatures $\{e_i\}_{i=1, \cdots r}$, which can be used to obtain the distributions $\{ \mu_{\theta}(e_i)\}_{i=1,\cdots,r}$, and participation factors $\{p_i\}_{i=1,\cdots,r}$. If the first event signature is perturbed by a small amount $\Delta e_1$, then the log-likelihood of the full event tensor changes by the following expression:
\begin{align*}
    \Delta \mathbb{P}(x) =  log~ \mathbb{P}(e_1 + \Delta e_1) - log~ \mathbb{P}(e_1) \\
    + \sum_{i=1}^N \left(log~ \mathbb{P}(p^i_1(e_1 + \Delta e_1) | e_1 + \Delta e_1) - log~ \mathbb{P}(p^i(e_1) | e_1) \right),
\end{align*}
where the index $i$ runs over the sampled participation factors. The first two terms only depends on the perturbed event signature and the latter two depend only on the adaptability of the first participation factor map (the terms in the sum, which should be near zero for a well trained participation factor map). \nomenclature{$\mathbb{P}$}{Probability}

Independent Component Analysis (ICA) \cite{hyvarinen1999survey} seems to fit the tensor decomposition method best given this latter algorithmic requirement. In ICA, the independent vector components can be interpreted as participation factors and view the mixing matrix as event signatures. Unfortunately, ICA fails to converge on about $10\%$ of the data tensors. As such, uncorrelated (orthogonal) participation factors instead of truly independent ones will be used in this study. A variant of Singular Value Decomposition (SVD) was applicable.
While SVD does yield orthogonal event signatures, this property is not required since distinct event signatures may overlap in time while still being statistically independent in the PMU space.
Removing this requirement means that it is appropriate to either perturb event signatures or replace them entirely - so long as any changes made preserve the orthogonality of the participation factors.

\subsubsection{The QR Re-Orthogonalization Trick} The participation factors can always be re-orthogonalized with a $QR$ factorization. To be specific, suppose there is an existing Event-Participation decomposition $X = PE^T$. If $E$ is changed to $\tilde{E}$, then a change in participation factor is induced
resulting in $\tilde{P}$ (to maintain the equality). If a $QR$ decomposition is performed on $\tilde{P}$, then $X = QR\tilde{E}^T$.
Denoting $Q$ as the new set of participation factors and $R\tilde{E}^T$ as the new set of event signatures, then we have a new orthogonal-satisfying event participation decomposition. This does mix the event signatures. However, since $R$ is upper-triangular, some subspace uniqueness can be kept by placing signatures that should be preserved at the bottom of the event-signature matrix.
\subsection{A Full Description of the Decomposition Steps}
In this subsection, the process of obtaining this event-participation decomposition from PMU data is described.
\subsubsection{Preprocessing}
The same standardization technique is applied on every data tensor in the PMU dataset.
The technique used is similar to the z-score standardization used in machine learning but differs in that the temporal means and standard deviations (which are different for each PMU) are calculated.
If the dimensions of the tensors are arranged as (PQVF, PMU index, time)\nomenclature{$P$}{Real Power}\nomenclature{$Q$}{Reactive Power} \nomenclature{$V$}{Voltage Magnitude} \nomenclature{$F$}{Frequency}, then the statistics are taken over the final axis.
In contrast to typical machine learning standardization techniques, this mean over the batch of data (i.e., over different events) is not taken.
This is because removing the mean over the time axis is critical to the privacy of the PMUs. PMU data, in raw form, typically has each PMU at a unique level of data. For example, one PMU may have a voltage offset of $230 kV$ and another an offset of $345 kV$. This property would be the same across all data tensors and make every PMU uniquely identifiable through their temporal statistics.

\subsubsection{Inter-event and Intra-event components}

\begin{figure}[h]
    \centering
    \begin{subfigure}{0.49\linewidth}
        \includegraphics[width=\linewidth]{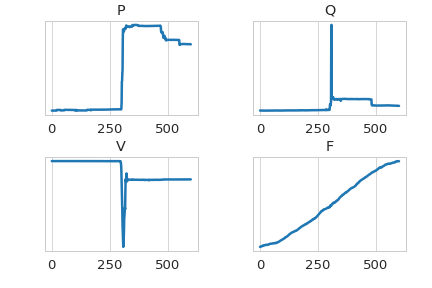}
    \end{subfigure}
    \begin{subfigure}{0.49\linewidth}
        \includegraphics[width=\linewidth]{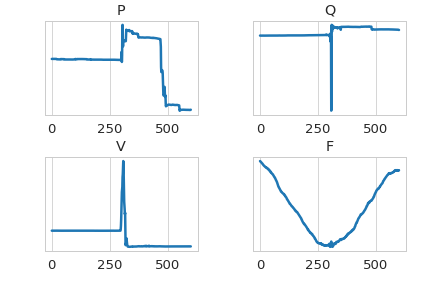}
    \end{subfigure}
    \caption{Top two inter-event signatures for voltage events.
    }
    \label{fig:inter-event-voltage-top2}
\end{figure}

\vspace{-0.3cm}

\begin{figure}[h]
    \centering
    \begin{subfigure}{0.49\linewidth}
        \includegraphics[width=\linewidth]{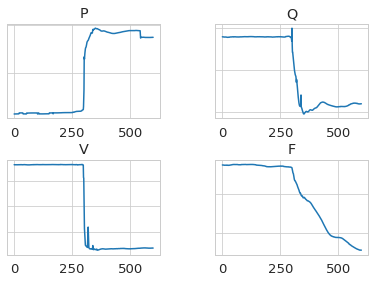}
    \end{subfigure}
    \begin{subfigure}{0.49\linewidth}
        \includegraphics[width=\linewidth]{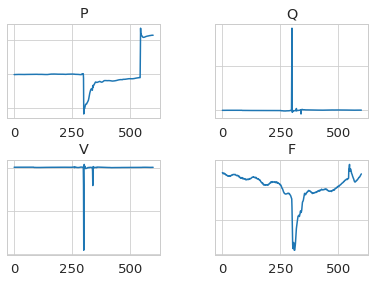}
    \end{subfigure}
    \caption{Top two inter-event signatures for frequency events.
    }
    \label{fig:inter-event-frequency-top2}
\end{figure}

Some dynamical behaviors are so fundamental to power system events that they end up repeating themselves, nearly in replica, across multiple events. These behaviors tend to lend themselves to simple verbal descriptions like ``frequency drops," ``voltage step changes," and ``spikes." A realistic set of generated data should include these behaviors as event-signature components.

To obtain these components, Principal Component Analysis (PCA) is performed on a sample of PMUs from every single event in the dataset. The obtained eigenvectors are the desired event signatures. However, since this PCA only used a sample of PMUs from each event, the coordinates returned by PCA do not yield a sufficient set of participation factors. Instead, projections onto these event signatures must be performed to obtain participation factors for every event tensor. Note that the factors obtained via this method are not necessarily orthogonal, so this step requires the first use of the $QR$ re-orthogonalization trick presented in the previous subsection. 

\begin{figure}[t]
    \centering
    \begin{subfigure}{0.49\linewidth}
        \includegraphics[width=\linewidth]{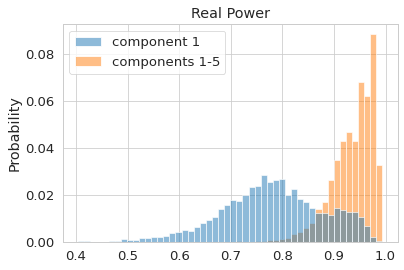}
    \end{subfigure}
    \begin{subfigure}{0.49\linewidth}
        \includegraphics[width=\linewidth]{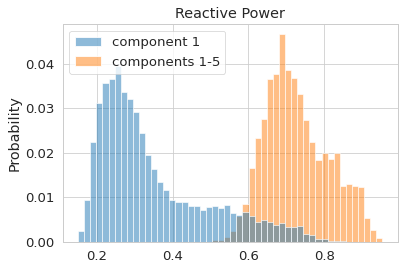}
    \end{subfigure}
    \begin{subfigure}{0.49\linewidth}
        \includegraphics[width=\linewidth]{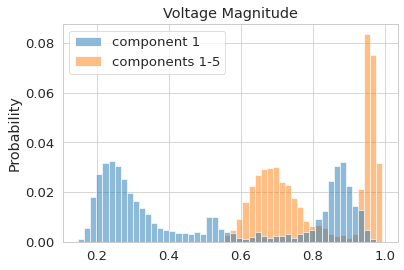}
    \end{subfigure}
    \begin{subfigure}{0.49\linewidth}
        \includegraphics[width=\linewidth]{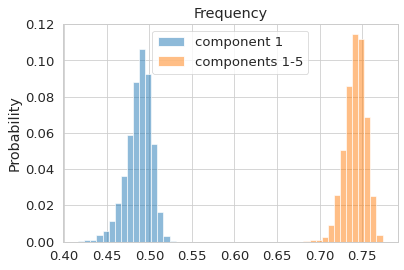}
    \end{subfigure}
    \caption{The full distribution of the explained variance ratios for the first event signature, and the cumulative explained ratio for all five event signatures}
    \label{fig:inter-event-var}
\end{figure}

The event signatures obtained this way are denoted as inter-event signatures. Due to space limitation, only the top two of the five inter-event signatures for voltage events learned from a random sample of 500 events are displayed in Figure \ref{fig:inter-event-voltage-top2} and those of frequency events in Figure \ref{fig:inter-event-frequency-top2}
The left subfigure is the first inter-event signature and the right subfigure is the second inter-event signature.
The full inter-event signatures can be found in Appendix A of the arXiv version of the paper.
To get an accurate estimate of the variance explained by these inter-event signatures, a bootstrap analysis of the explained variance ratio for each component is performed. The bootstrap was performed $5000$ times. Each bootstrap consisted of re-sampling $2000$ PMUs with replacement, calculating five inter-event components over those $2000$ PMUs, and then calculating the proportion of variance explained by these components. 
The full distribution of the explained variance ratios for the first event signature, and the cumulative explained ratio for all five event signatures, are plotted in Figure \ref{fig:inter-event-var}. It can seen that, over the large number of events that was gathered, real power and voltage magnitude data has significantly more globally similar behavior than reactive power and frequency data do.

\subsubsection{Intra-Event Signatures}
The inter-event signatures give us an event-participation decomposition, yielding a surprisingly decent approximation. However, they do not capture all subtleties of an event. 
To capture the intricacies of an event tensor, the inter-event approximation of that event is first taken. Again, this amounts to getting the inter-event signatures from all 
events and then performing a projection for the specific event that is being decomposed to get its inter-event participation factors.
This approximation is denoted as $X_{\text{inter}}$.
If the current event tensor under consideration is denoted $X$ \nomenclature{$X$}{Data Tensor}, then the residual $X_{\text{inter}} - \tilde{X}$ must be further decomposed.

For this final decomposition, a mix of sparse-PCA \cite{zou2006sparse} and standard SVD is used. The reason for using two types of decomposition methods is straightforward - some event dynamics do not last very long. So the signature of such dynamics should be zero most of the time. To ensure orthogonality of the participation factors,  
the sparse-PCA decomposition is performed to extract the top $5$ components.
Note that, with sparse-PCA, the resulting participation factors are not guaranteed to be orthogonal. 
Thus, the $QR$ re-orthogonalization trick is applied again for this set of extracted event signatures.
The resulting participation factors are denoted as $P_{\text{sparse}}$ and the event signatures as $E_{\text{sparse}}$.
Then an intermediate decomposition is formed to approximate $X_{\text{inter + sparse}} \approx \tilde{X} + P_{\text{sparse}}E_{\text{sparse}}$. 
Finally, the standard SVD is performed on the residual $X - X_{\text{inter + sparse}}$ to obtain the final $15$ event signatures. 

\subsection{Summary}
At the end of this process, a set of $25$ event signatures and their corresponding participation factors can be obtained for each event.
Five of these event signatures exist across all events, and the remaining $20$ are unique to each event. Of the $20$ unique signatures, $5$ consist of re-orthogonalized sparse event signatures, and the final $15$ consist of non-sparse entries. The event-participation pairs are statistically independent via the orthogonality of the participation factors - which can be viewed as independent and identically distributed samples from a distribution that depends on the event signature.

\section{Generating Participation Factors}\label{sec:Generator}


The deep learning model for generating participation factors is described in this section. Real-world dataset shows that a typical distribution of inter-event participation factors is multimodal, non-Gaussian in which the per-mode distribution varies by both shape and size. Thus, the model for generating participation factors should satisfy the following requirements.
\begin{enumerate}
    \item The model must generate samples from different distributions depending on the input event signature.
    \item The model must simultaneously scrutinize all data types (PQVF) and output samples for all data types.
    \item The model must be time-invariant - the same event translated by a small amount in time should yield the same set of participation factors.
    \item The model must output a multimodal distribution (whose modal positions and shapes depend on input signatures). 
\end{enumerate}

Due to the often subtle differences in event signatures, this first point requires a fairly expressive deep learning model. The model will thus need a relatively large number of parameters.

The description of the proposed model will proceed top-down.
The high-level overview of the model is first provided, which is follow by the description of its components.
Some of these components are also subdivided into a high and a lower-level structure, which will be explained following the top-down approach.
\subsection{Full Model Overview} \label{subsec:overview}
\begin{figure}[t]
    \centering
        \includegraphics[width=\linewidth]{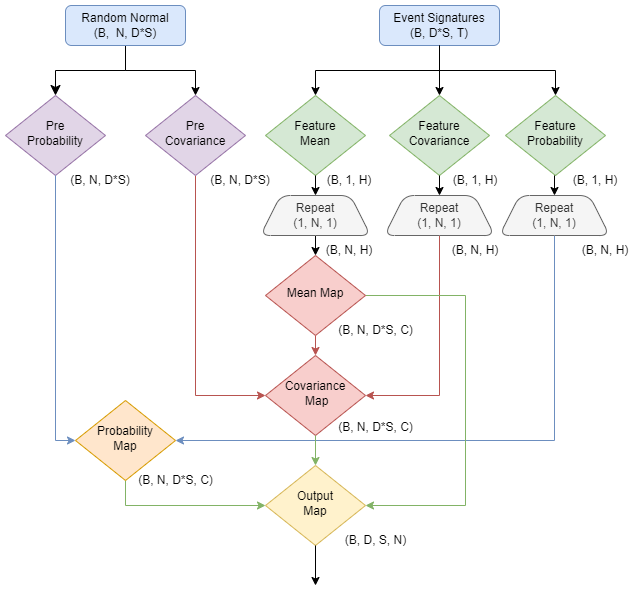}
    \caption{
    An overview of the generative model used to simulate event-dependent participation factors.
    }
    \label{fig:pfgen_over}
\end{figure}

A diagrammatic view of our full generative model is shown in Figure \ref{fig:pfgen_over}. 
The event signature is first shaped to $(B, D*S, T)$ where $B$ is the batch size\nomenclature{$B$}{Batch Size}, $D$ \nomenclature{$D$}{Measurement Channel Size (set at $4$ for PQVF)} is $4$ (for PQVF), $S$ is the number of coinciding event signatures\nomenclature{$S$}{Number of Coinciding Event Signatures}, and $T$ is the temporal horizon of the data. The event signatures for each datatype are stacked into one axis.
The $i$-th index in this axis corresponds to the $\text{floor}(i/4)$-th event signature of the $i (\text{mod}) 4$-th datatype (ordered by $P=0$, $Q=1$, $V=2$, $F=3$).

This stacked tensor goes through three feature extraction maps, denoted as ``Feature Mean", ``Feature Covariance", and ``Feature Probability". 
These maps result in $H$ hidden representation variables for each data point in the batch, encapsulating essential details in the 
event signatures. Each of these variables is repeated $N$ times so that the same set of hidden features can be used for every generated PMU. Simultaneously, a 
normal random variable of shape $(B, N, D*S)$ is generated.
This tensor goes through two ``Pre-Maps" denoted as ``Pre Probability" and ``Pre Covariance." These maps capture global non-Gaussian behavior amongst modes. 
The outputs of these maps have the same shape as their inputs.

The output of ``Feature Mean" then goes through a map denoted as ``Mean Map", and the output of that map (as well as the output of ``Pre Covariance") goes through a map called ``Covariance Map". 
The outputs of these maps have shape $(B, N, D*S, C)$. Here, a new dimension has been added, denoted $C$. This dimension captures any multi-modal behavior present in the distributions. The output of ``Mean Map" represents the locations of these modes and only uses the ``hidden features" dimension of its input. Since these features are the same for every generated PMU, the mode locations are also the same for every PMU. The ``Covariance Map", in contrast, allows for differing values for each PMU but forces the average value over those PMUs to be zero.

Simultaneously, the output of the ``Feature Probability" map and the ``Pre Probability" map feed into the ``Probability Map", which probabilistically assigns each of the generated PMUs to one of the modes. Finally, the outputs of the probability map, mean map, and covariance map are combined as shown in (1):
\begin{equation}
    \text{out}[i, j, s] = \sum_c p[i, j, s, c] \cdot (\mu[i, j, s, c] + \Sigma[i, j, s, c]),
\end{equation}
where $p$ refers to the output of the ``probability map", $\mu$ to the output of the ``Mean Map", and $\Sigma$ to the output of the ``Covariance Map". 

\subsection{Feature Extraction Maps}
\begin{figure}[t]
    \centering
    \includegraphics[width=\linewidth]{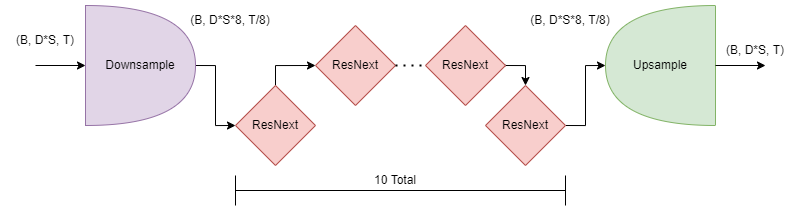}
    \caption{The identical structure of each feature map.}
    \label{fig:cascade}
\end{figure}

The three feature extraction maps, ``Feature Mean", "Feature Covariance", and "Feature Probability", all share a similar structure. These networks are illustrated in Figure \ref{fig:cascade}. They are all given by a cascading convolutional neural network with a ResNext block serving as its inner channel. 
The following two subsubsections detail these fundamental components.

\subsubsection{Cascaded Convolutional Network}
We adopt the cascading convolution network framework because it is very effective in performing reconstruction of high-quality dataset with low signal-to-noise ratio and avoids the overfitting problems of a basic convolution neural network. A cascading convolutional neural network is a deep neural network consisting of three units: a down-sampling cascade unit, a fundamental block, and an up-sampling cascade unit. 

The down-sampling cascade unit uses several layers of one-dimensional convolutional filters of size $3$ with a stride of $2$ and padding of $1$, which guarantees its output (in the time dimension) is exactly half the size of its input. 
However, the overall size of the input data is preserved at each layer by doubling the number of channels. In pmuBAGE, we used $3$ down-sampling layers. Each layer is followed by a one-dimensional instance norm, and a Scaled Exponential Linear Unit (SELU) activation function.

The fundamental block used for the feature maps in pmuBAGE is a $5$-channel ResNext block described in the following subsubsection. This block is repeated $10$ times sequentially (each with differing parameters). The up-sampling cascade unit mirrors the down-sampling unit exactly. However, it uses transposed convolutional filters to \textit{double} the temporal length of each sequential input while \textit{halving} the number of channels. They too are followed by a one-dimensional instance norm, and a SELU activation function.

After the up-sampling cascade unit, an adaptive average pooling layer is used to receive $H$ hidden units. These units capture the relevant information from the event signatures for each map. In pmuBAGE, $H=1000$.

\subsubsection{ResNext Convolutional Blocks}

The five-channel ResNext block sends the input through five parallel channels - the outputs are summed together. One of these channels does not change the input at all. The other four are identical in structure. These identical channels go through the following sequence of transformations. The first transformation is a length-$1$ convolution filter. 
The number of channels reduces to $4$ upon this transformation. The subsequent transformations are a BatchNorm layer followed by a SELU activation layer. A length-$3$ convolutional filter then follows a length-1 reflection pad (to keep the length of the output unchanged), which is then followed by another SELU layer and yet another BatchNorm. All of the convolutional filters in this block use a stride of $1$ and no padding (other than the explicit reflection pad mentioned above).

\subsection{Pre-Maps}

The pre-maps, ``Pre Probability" and ``Pre Covariance" both take the randomly thrown normal random variable $z$ \nomenclature{$z$}{Random Normal Sample Tensor} as input. The maps are intended to capture any global (across modes and events) non-Gaussian behavior. These maps have an identical structure given by a residual sequential layer of a size-preserving Dense Unit, a SELU activation, and another size-preserving Dense Unit. That is, the output of each of these maps is given by the following equation:
\begin{equation}
    z \mapsto z + (\text{Dense}_2 \circ \text{SELU} \circ \text{Dense}_1)(z)
\end{equation}
\subsection{Discrete Probability Map}
The map labeled ``Probability Map" assigns each generated PMU to a mode of the output distribution. This map first concatenates its two inputs into one tensor with shape $(B, N, D*S + H)$. It then takes this concatenated input through a Dense shape-preserving layer followed by a SELU activation. It then sends the activation output through a Dense expansive layer - this time mapping the $D*S + H$ to $D*S*H$ units. This output is then reshaped to $(B, N, D*S, H)$ before being mapped through another Dense layer - this time contractive, mapping the $H$ hidden units to $C$ units (where $C$ is the number of categories). This is followed by a final SELU activation function and one last (shape preserving) Dense layer. The output of this last dense layer represents the (pre-normalization) probability mass function over the categories. Samples can be drawn from this distribution by applying softmax and performing a discrete sampling procedure.
However, gradients can not be send back through the discrete sampling procedure. 
This means that it is infeasible to learn the previous parameters that define the probability of each mode as a function of the input.
Instead, an approximation of this sampling procedure known as the Gumbel-Softmax Function can be used, which is detailed in reference \cite{jang2016categorical}.

\subsection{Mean and Covariance Maps}
The mean map is simply a sequence of a Dense layer (mapping the $H$ hidden input layers to $S*D*C$ output layers), a SELU activation layer, and another Dense layer (this time size-preserving). The output of the final dense layer is reshaped to $(B, N, D*S, C)$. The covariance map first undoes the final reshape of the mean map before concatenating the resulting tensor with the outputs of the ``Feature Covariance" Map and the ``Pre Covariance" Map along the final dimension. This results in an input tensor of shape $(B, N, D*S*C + H + D*S)$. This tensor is then sent through a Dense layer (reducing the $D*S*C + H + D*S$ hidden units to $D*S*C$ new hidden units), a SELU activation layer, and a final (size-preserving) Dense Layer. The output of the final dense layer of this map is also reshaped to $(B, N, D*S, C)$.

\subsection{Loss Functions}\label{subsec:losses}
To train this model, three categories of loss functions are used.
The first is the standard Generative Adversarial Network (GAN) loss function. 
The second category consists of several feature matching loss functions. The statistics of the generated participation factors (for a given set of event signatures) are compared to the actual participation factors corresponding to that set of signatures. The third category of loss functions is a new loss function that is invented for this work, denoted ``quantile loss." Explicit descriptions of all of these losses are provided in the following subsubsections.

\subsubsection{GAN Loss and Discriminator Design}
The mean squared error criterion is selected as the GAN loss.
For this loss, the generator outputs $G(x)$ \nomenclature{$x$}{Tensor of Event Signatures} are sent into a discriminator function $D(x, \cdot)$, which scrutinizes its second input 
and returns a value representing the probability that it thinks the second input was real. The following equation explicitly gives the generator loss with respect to its discriminator output:
\begin{equation}
    \mathcal{L}_{G_{\text{disc}}} = \mathbb{E}_{\text{batch}}[\| D(x, G(x)) - 1\|^2]
\end{equation}
This error is low when the generated points trick the discriminator into outputting a value near $1$ (indicating that the discriminator thinks the PMU data is likely to be real).

The discriminator is itself trained so as to not be tricked by the generator, using the following loss function:
\begin{equation}
    \mathcal{L}_{D} = \mathbb{E}_{\text{batch}}[\|D(x, G(x)) - 0\|^2] + \mathbb{E}_{\text{batch}}[\|D(x, p(x)) - 1\|^2]
\end{equation}
where $p(x)$ are the real participation factors corresponding to the event signatures in $x$.

The discriminator consists of two maps. The first is a Cascaded Convolutional Neural Network exactly equivalent 
to the feature extraction maps of the generator. This Cascade Network takes in the event signatures (the first argument of $D$) and returns a set of $H$ extracted hidden features. These $H$ features are concatenated to the second argument of $D$ to give a tensor of shape $(B, S*D, H+N)$. This larger tensor is then sent through a size-preserving Dense Layer followed by a SELU activation, and then a final Dense layer mapping the $H+N$ units to $1$ final unit. Often a sigmoid activation is placed on this final unit to ensure stability (at the risk of vanishing gradients). However, it is discovered that this is unnecessary during our training stages.
The output of this unit was always bound to the interval $[0, 1]$ without the sigmoid intervention.

\subsubsection{Feature Match Losses} In addition to this GAN loss, several ``feature matching" losses were included. Feature matching losses consist of taking expectations of several functions of the generated data and comparing them to the expectations of those same functions applied to the actual data (via absolute differences). For pmuBAGE, some of these expectations were taken along the PMU-axis. This gives the loss functions more event-signature-dependent resolution than the standard practice of taking batch-axis expectations. However, for some functions, the proposed model found learning at this resolution difficult
, and so for those functions, expectations were taken along both the PMU-axis and the batch-axis. 

The functions that used for this purpose are as follows:
\begin{align*}
    f_1(x) &= x & f_2(x) &= \text{min}_{\text{PMU-axis}}(x) \\
    f_3(x) &= \text{max}_{\text{PMU-axis}}(x) & f_4(x) &= \sigma(x * 2 / (f_3(x) - f_2(x))) \\
    f_5(x) &= f_4(-x) & f_6(x) &= relu(x) \\
    f_7(x) &= relu(-x) & f_8(x) &= relu(x)^2 \\
    f_9(x) &= relu(-x)^2 & f_{10}(x) &= x^3,
\end{align*}
where $\sigma$ refers to the sigmoid function. 
All functions appearing with powers greater than $1$ were centralized before taking their expectations.
The first $9$ of these functions used only the PMU-axis for calculating their expectations. The $10^{th}$ function needed both the PMU-axis and the batch-axis. The inclusion of the first three functions is straightforward. The fourth and fifth functions attempt to capture the proportion of PMUs with participation factors greater than zero and less than zero, respectively. The sixth, seventh, eighth, and ninth functions attempt to capture the first and second-order statistics of the PMUs restricted to positive or negative participation factors.
The $10^{th}$ function attempts to capture any left/right skew that may occur in the dataset. Finally, a feature map loss function in the form of the covariance matrix (calculated across the PMU axis) of every pair of Event Signatures present is included.
Note that these covariances will differ between real and fake data since they are calculated using the participation factors. 

\subsubsection{Quantile Losses} The final category of loss function is new.
To calculate this function, let us first find the quantiles (refined in steps $\Delta Q$\nomenclature{$\Delta Q$}{Quantile Loss Step Size}) of the real PMU data (along the PMU axis). Label these quantiles as $q_i$. \nomenclature{q}{Quantile Locations} For example, if $\Delta Q$ is $0.1$, the interval $(-\infty, q_1)$ contains $10\%$ of the real PMU participation factors. In general, the proportion of real points with participation factor exceeding $q_i$ is $1 - i\Delta Q$. This property needs to be maintained in the generated dataset for each $i$.
The proportion of samples exceeding the value $q_i$ is equal to the mean value of the Heaviside function (with step change occurring at $q_i$). Unfortunately, the Heaviside function's gradient is zero, so implementing this would immediately guarantee vanishing gradients for this term. To avoid the problem, the Heaviside function is approximated with a scaled sigmoid function $\sigma(a_i x)$.
Scaling the sigmoid function allows us to control the width of the window in which the gradient is nonzero. A tiny window (i.e., a very high $a_i$) will only affect the location of PMUs very near the quantile value under consideration. In totality, the value ${\mathbb{E}_{\text{generated}}[\sigma(a_i \cdot (x-q_i))]}$ is compared against
$1-i \Delta Q$ (with comparison formed via absolute difference).
For pmuBAGE, the value of $\Delta Q$ was set to $1 / 25$, and $a_i$ was set to $\frac{2}{q_{i+1} - q_i}$.

\subsection{Simulating Inter-Event Signatures}\label{subsec:inter}
Learning the participation factor distributions for the inter-event signatures presents a new problem: the signatures are the same for every event but have different distributions. Thus, if these event signatures were used as inputs without context,
then the map from these signatures to the correct distribution would not be a function at all - it would have multiple valid outputs.
Technically, this is not exactly the situation
because every input to our model contains the other ``fine detail signatures" (i.e., the sparse signatures and intra-event signatures). However, this would mean that learning the correct participation factor distribution for the inter-event signatures boils down to memorizing fine detail signatures. This seems non-ideal since the event signatures should interact as little as possible.
In particular, the interaction between the fine-detail signatures and the inter-event signatures should be minimized since
the specific properties of a given event should not affect the large-scale properties of the output dataset.

The simplicity of the inter-event signatures gives their participation factors properties that can be easily mimicked without the need of deep learning. 
The critical insight into our inter-event participation factor generator comes from using sorted
participation factors. That is, for each event, each inter-event signature, and each data type (PQVF), the corresponding PMUs are sorted in descending order.
Thus, for each event, data type, and inter-event signature, the first listed PMU corresponds to the highest participating PMU, the second for the second-highest PMU, and so on. Notably, the highest participating PMU for one data type is not necessarily the same PMU as another. Thus a single instance of this rearranged instance does not necessarily correspond to any single PMU. 

When this transformation is performed, two useful properties emerge. The first is that, for every single inter-event signature, a single PCA component explains $90\%$ of the transformed dataset's variance. The second has to do with the coordinates obtained via projecting the dataset onto this single principal eigenvector. As it turns out, different data types of this double transformed dataset are in near perfect linear relation. 
Furthermore, the distribution of real power data under this transformation is close to a perfect Gaussian. Thus, to simulate inter-event participation factors, we only need to throw a set of random normal variables of appropriate mean/variance
, apply regression coefficients to obtain the corresponding data for reactive power, voltage magnitude, and frequency, and then multiply the result by the single extracted PCA eigenvector. This process gives us a distribution of inter-event signatures for each data-type that closely resembles that of the original PMU dataset. In fact, a Kolmogorov-Smirnov test yields a test statistic of $0.13$ for the 
real power distribution ($p=36.8\%$), $0.16$ for the reactive power distribution ($p=15.5\%$), $0.90$ for the voltage magnitude dataset ($p=81.5\%$), and $0.13$ for the frequency dataset ($p=36.7\%$). Under these values, the null hypothesis that the two samples come from the same distribution won't be rejected.

\vspace{-0.3cm}

\section{Numerical Results}

\subsection{Training Dataset}

Our training dataset consists of $PQVF$ data for $620$ voltage events and $84$ frequency events. Data was collected from $180$ PMUs across the Eastern Interconnection of the U.S. Transmission Grid at a sampling rate of $30 Hz$. The original synchrophasor data consists of voltage magnitude, voltage angle, current magnitude, current angle, and frequency decomposed as symmetrical components. Some synchrophasors reported all three symmetric components for each data type (positive sequence, negative sequence, and zero sequence), but most only reported the positive sequence component. Thus only positive sequence values were used to train our model. $PQVF$ data was derived from the values mentioned above.

Each event in our training dataset consists of $20$ seconds ($600$ samples) of data, with the event's start placed at the $10$ second mark ($300$-th sample). The exact event timing was not always reported accurately, so expert analysis either affirmed or adjusted these start times by a few fractions of a second. Frequent missing data samples appeared during event windows. PMUs that exhibited such values were removed from the dataset, but only for the event where they had missing data. Due to space limitation, synthetic PMU data for voltage and frequency events and sub-event categories are not shown in this manuscript, but can be found in Appendix B and C of the arXiv version of the paper.

\vspace{-0.2cm}

\subsection{Training Details}

The parameters of our model were trained for $500$ epochs with a batch size of $50$ events per batch. The Adam optimizer was employed, and an $l_2$ weight regularization term with a coefficient of $0.25$ was applied. Generative Adversarial Networks are known to have notorious training difficulties. The most common difficulty occurs when the discriminator converges too quickly. This forces the discriminator's feedback to the generator to have vanishing gradients. To avoid this, the learning rates set for the discriminator's parameters were significantly lower than that of all other parameters. The learning rate of discriminator parameters is set to $1\times10^{-5}$ while all other parameters used a learning rate of $1\times 10^{-3}$.
Training this model proceeds smoothly.
Both loss functions dip to low values quickly during training, followed by a somewhat slow-moving period in which both curves converge to their near-optimal values. 
Loss curves for those feature matching losses are similarly stable but converge much more quickly.

\vspace{-0.2cm}

\subsection{Similarity Analysis}
While it is more desirable that the resulting generated events to be similar to the training set, they should not be identical.
To ensure some level of non-similarity to the original dataset, a correlation analysis is performed.
To do so, the datasets are partitioned 
in two different ways - first, by Event ID, and second, by PMU ID. For the Event ID partition, the full tensor inner product of the synthetic event tensor and the measured event tensor are computed for each event.
For the PMU ID partition, the full tensor inner product for the PMU after concatenating all events that the PMU is in is calculated for each PMU.
The Event ID partition will help identify if any of the synthetic events in the dataset are too similar to the corresponding measured event (the event with the same set of event signatures as the synthesized tensor). The PMU ID partition will help identify if any PMUs have been compromised by the synthetic dataset. 

\begin{figure}[]
    \centering
    \begin{subfigure}{0.49\linewidth}
        \includegraphics[width=\linewidth]{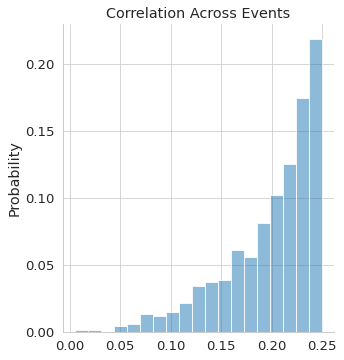}
    \end{subfigure}
    \begin{subfigure}{0.49\linewidth}
        \includegraphics[width=\linewidth]{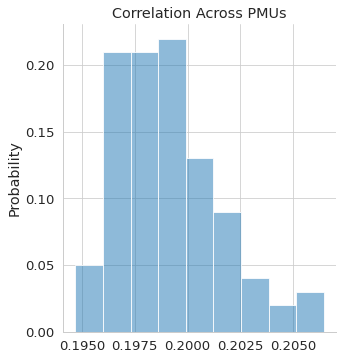}
    \end{subfigure}
    \caption{Correlation histograms between the synthetic and measured PMU datasets partitioned by event and PMU ID.
    }
    \label{fig:corr}
\end{figure}

Two histograms of correlation values are plotted in Figure \ref{fig:corr}. On the left, each datapoint used in calculating the histogram was an Event ID partitioned correlation value. There is a peak at correlation $0.25$, but this is the maximum of such correlation. As such, it is deemed that no events are too similar to their original incarnation. On the right of Figure \ref{fig:corr}, each datapoint used in calculating the histogram was a PMU ID partitioned correlation value. No correlation exceeds $0.21$, so it is deemed that no PMU has been compromised.

\vspace{-0.2cm}

\subsection{Inception-Like Scoring}
One of the most well-accepted ways to analyze the quality of generated data samples is through a measure called the ``Inception Score." A universally agreed-upon third-party classification model, called the Inception Model, is trained on a set of measured data to calculate an inception score. Generated samples are then fed into the Inception Model, and the classification accuracy of the Inception Model on those generated images is reported. A high classification accuracy indicates that the generated images are of high quality.

Unfortunately, there is no standard Inception Model for the generated event samples.
Thus, the quality of these event samples will be assess with an approximation of the Inception Score.
To do this, a relatively standard ResNext \cite{Xie2016} Model (equivalent to each of the ``Feature Extraction Maps" presented above) is adopted and trained to classify event types with class labels ``Frequency", and ``Voltage".
Training of this classification model was performed over $200$ epochs with a batch size of $50$ using the Binary Cross Entropy loss function. Training was performed with the Adam optimizer with a learning rate of $1\times 10^{-3}$. Training proceeded smoothly with near monotonically-decreasing loss.

Training was performed twice - once with the synthetic data and once with the measured data. Each training data point was 20 second event window - either synthetic or real depending on the experiment - unmodified from its original form. Each trained model is then tested on the synthetic and measured PMU data, which yields four different train-test scenarios:
\begin{enumerate}
    \item Train on Synthetic, test on Synthetic
    \item Train on Synthetic, test on Measured
    \item Train on Measured, test on Synthetic
    \item Train on Measured, test on Measured
\end{enumerate}

Suppose the cross scores (Synthetic-Measured, Measured-Synthetic) are not significantly degraded from the self scores (Synthetic-Synthetic, Measured-Measured). In that case, the quality of the generated samples are deemed to be high.
In terms of interpretation of these scores, the Synthetic-Measured score indicates how well a model trained on these generated events will perform in practice. In contrast, the Measured-Synthetic score indicates how well a model trained on measured data would benchmark against pmuBAGE.

Results are reported in Table \ref{tab:inception}. Since the measured and synthetic datasets are imbalanced (with a ratio of about ${7:1}$ in favor of voltage events), the F1 and F2 scores are reported alongside classification accuracy.
In training each classification scenario, each training loss for voltage events is multiplied by $1/7$ so that their back-propagation gradient descents would be weighted equally to those of frequency events.

\begin{table}[h]
\caption{Self and Cross Scores Across Different Scenarios
}
    \centering
    \def\arraystretch{1.5}
    \begin{tabular}{l | l  l  l}
    \hline
         Training-Testing & Acc. & F1 & F2 \\ \hline
         Synthetic-Synthetic & $99.9\%$ & $94.3\%$ & $93.3\%$ \\
         \textbf{Synthetic-Measured} & $\mathbf{94.3\%}$ & $\mathbf{94.2\%}$ & $\mathbf{92.8\%}$ \\
         Measured-Measured  & $99.8\%$& $94.4\%$ &  $91.2\%$ \\
         \textbf{Measured-Synthetic} & $\mathbf{93.2\%}$ &  $\mathbf{94.3\%}$ & $\mathbf{92.7\%}$ \\ \hline
    \end{tabular}
    \label{tab:inception}
\end{table}

There appears to be no significant degradation in $F1$ or $F2$ scores in cross-comparisons compared to self comparisons, but there is a notable drop in accuracy. However, this drop is less than $7\%$, and all values are still greater than $90\%$. However, since the dataset is very imbalanced for this problem, the $F$ scores are more relevant since the accuracy measure is far less sensitive to the correct identification of frequency events than they are to voltage events. 

\section{Conclusion}
The synthesized PMU dataset created in this work, pmuBAGE, is highly realistic to the human eye and does not significantly degrade important training evaluation metrics compared to the measured PMU dataset from which it is trained. Thus, pmuBAGE may serve the scientific community as a standard benchmarking tool for algorithms which need PMU dynamic event data to train and validate on. Researchers should be aware of some pitfalls of the synthesized dataset. 
First, there is a lack of realistic banding for frequency data - that is, the tendency for several PMUs to have the same frequency behavior, especially during a frequency event. Instead, 
pmuBAGE tends to have PMUs act similarly in frequency data but with a slight spread in their exact values. Second, there is a small ``wrong direction" effect in voltage events voltage data. Actual voltage data undergoing a voltage event will primarily have PMUs drop the voltage values, while pmuBAGE has a small amount of PMUs raise their values very slightly. Third, pmuBAGE occasionally displays inter-area oscillations that do not dampen as quickly as actual data; however, the difference is relatively small. 
While these problems are expected to be eliminated in future version of this dataset, the current incarnation is sufficiently realistic to serve its purpose to the community - as a training, validating, and benchmarking dataset.


\printbibliography

\clearpage

\appendices

\section{Full Inter-event Signatures}

\begin{figure}[h]
    \centering
    \begin{subfigure}{0.49\linewidth}
        \includegraphics[width=\linewidth]{figures/InterEvents/0.png}
    \end{subfigure}
    \begin{subfigure}{0.49\linewidth}
        \includegraphics[width=\linewidth]{figures/InterEvents/1.png}
    \end{subfigure}
    \begin{subfigure}{0.49\linewidth}
        \includegraphics[width=\linewidth]{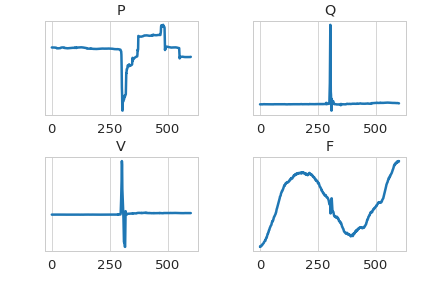}
    \end{subfigure}
    \begin{subfigure}{0.49\linewidth}
        \includegraphics[width=\linewidth]{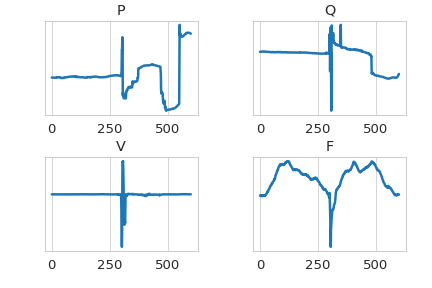}
    \end{subfigure}
    \centering
    \begin{subfigure}{0.49\linewidth}
        \includegraphics[width=\linewidth]{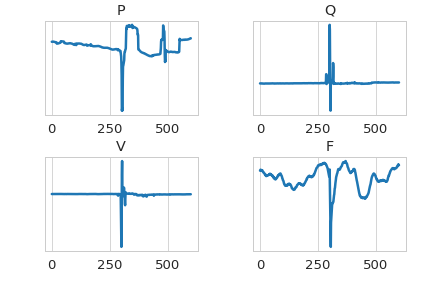}
    \end{subfigure}
    \caption{Full inter-event signatures for voltage events.
    }
    \label{fig:inter-event-voltage}
\end{figure}

\begin{figure}[h!]
    \centering
    \begin{subfigure}{0.49\linewidth}
        \includegraphics[width=\linewidth]{figures/InterEvents/f0.png}
    \end{subfigure}
    \begin{subfigure}{0.49\linewidth}
        \includegraphics[width=\linewidth]{figures/InterEvents/f1.png}
    \end{subfigure}
    \begin{subfigure}{0.49\linewidth}
        \includegraphics[width=\linewidth]{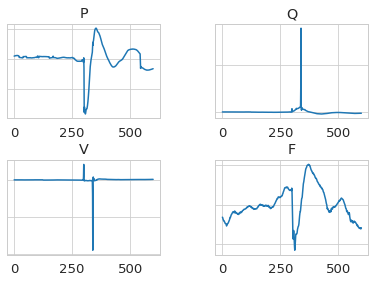}
    \end{subfigure}
    \begin{subfigure}{0.49\linewidth}
        \includegraphics[width=\linewidth]{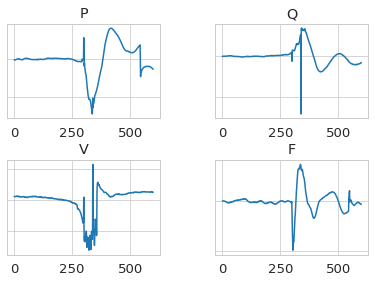}
    \end{subfigure}
    \centering
    \begin{subfigure}{0.49\linewidth}
        \includegraphics[width=\linewidth]{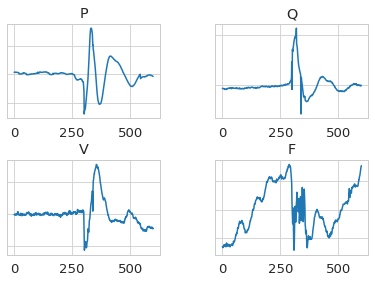}
    \end{subfigure}
    \caption{Full inter-event signatures for frequency events.
    }
    \label{fig:inter-event-frequency}
\end{figure}

All five inter-event signatures for voltage events 
are displayed in Figure \ref{fig:inter-event-voltage}, and those of frequency events in Figure \ref{fig:inter-event-frequency}. The top left subfigure is the first inter-event signature, the top right is the second, the bottom left is the third, the bottom right is the fourth, and the bottom one is the fifth.

\section{Actual PMU Data versus Synthetic PMU Data}

\begin{figure}[h]
    \centering
    \begin{subfigure}{0.49\linewidth}
        \centering
        \textbf{pmuBAGE}\par\medskip
        \includegraphics[width=\linewidth]{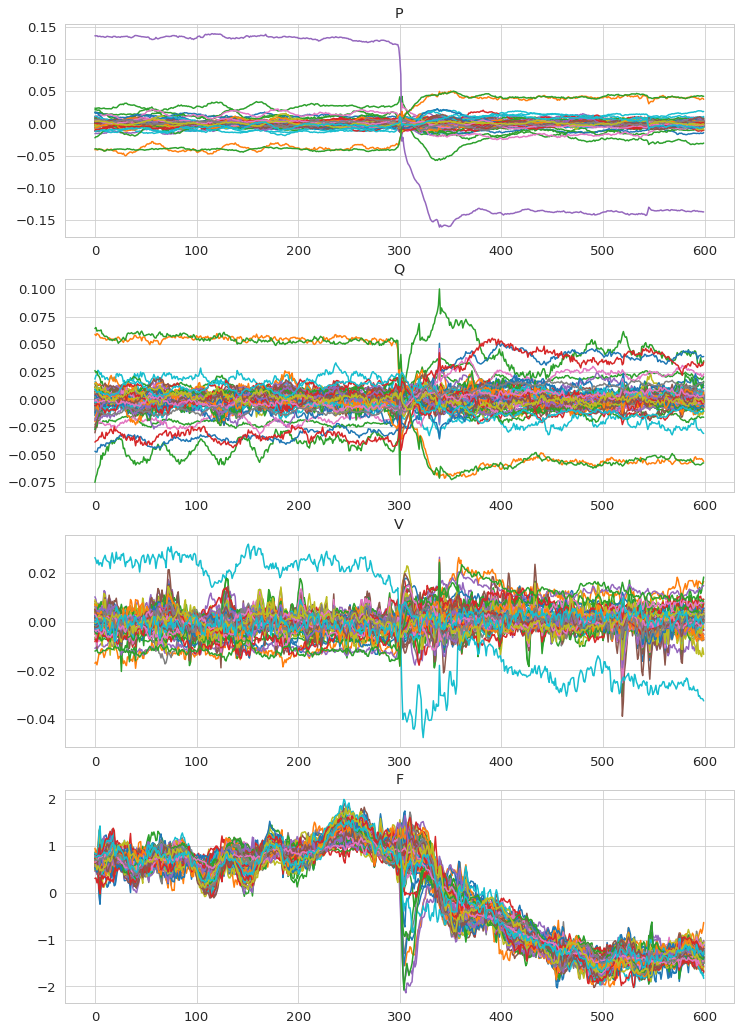}
    \end{subfigure}
    \begin{subfigure}{0.49\linewidth}
        \centering
        \textbf{Actual Data}\par\medskip
        \includegraphics[width=\linewidth]{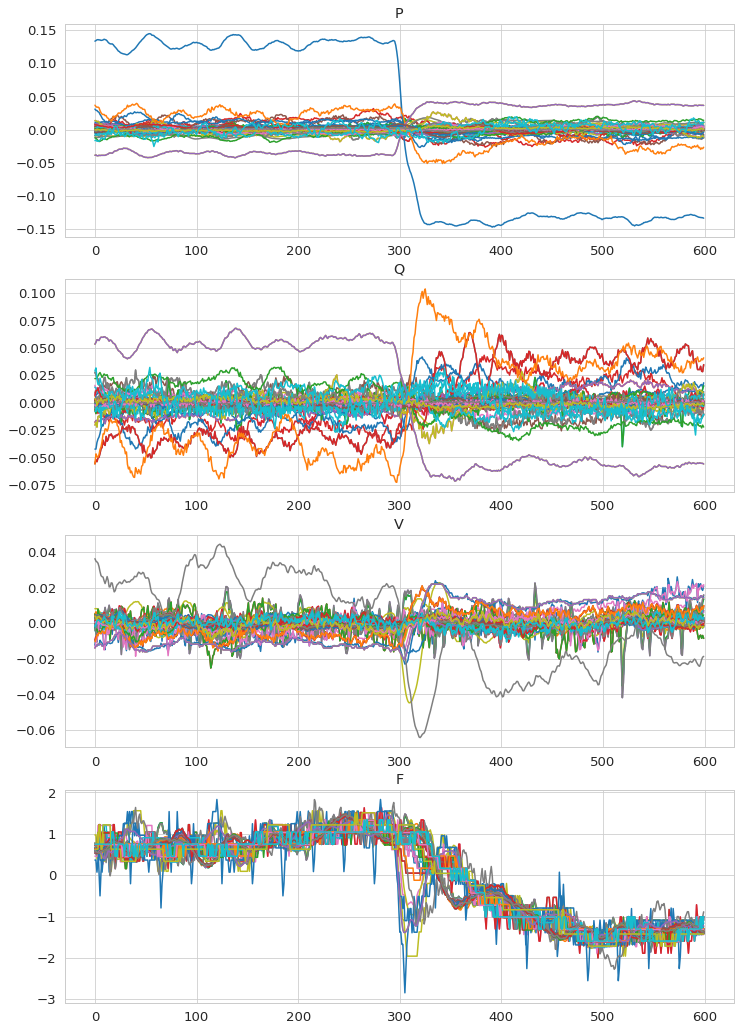}
    \end{subfigure}
    \caption{ A generated versus real frequency event.
    } 
    \label{fig:fevent_display}
\end{figure}

\begin{figure}[h]
    \centering
    \begin{subfigure}{0.49\linewidth}
        \centering
        \textbf{pmuBAGE}\par\medskip
        \includegraphics[width=\linewidth]{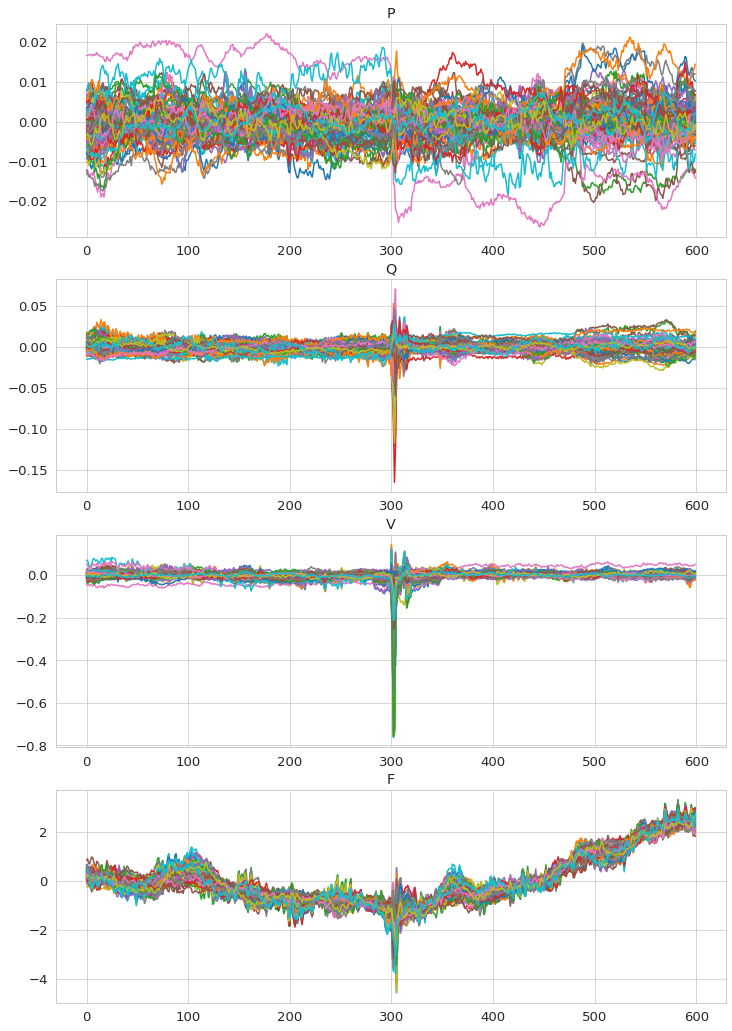}
    \end{subfigure}
    \begin{subfigure}{0.49\linewidth}
        \centering
        \textbf{Actual Data}\par\medskip
        \includegraphics[width=\linewidth]{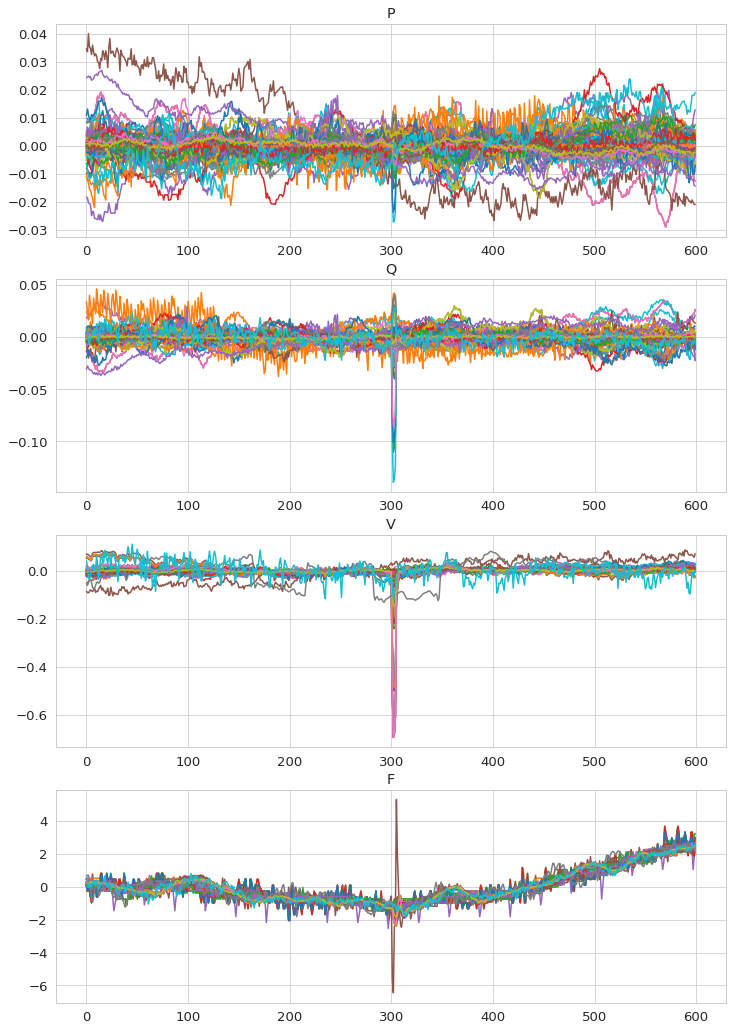}
    \end{subfigure}
    
    \caption{A generated versus real voltage event.
    } 
    \label{fig:vevent_display}
\end{figure}

Two example events are displayed in Figures \ref{fig:fevent_display} and \ref{fig:vevent_display}. The left subfigures are generated events in pmuBAGE and the right subfigures are real events of similar type.
Different colors represent different PMUs (100 signals). The interval between two time indices is 1 / 30 seconds. The full window corresponds to $20$ seconds of data. The presented data is scaled to per unit values and each time series is shifted to have a mean of zero. The four sub-figures represent real power, reactive power, voltage magnitude, and frequency from top to bottom. 

\vspace{-0.45cm}

\vspace{-0.6cm}

\section{Synthetic PMU Data for Sub-event Categories}

\begin{figure*}[h]
    \centering
    
    \begin{subfigure}{0.33\linewidth}
        \includegraphics[width=\linewidth]{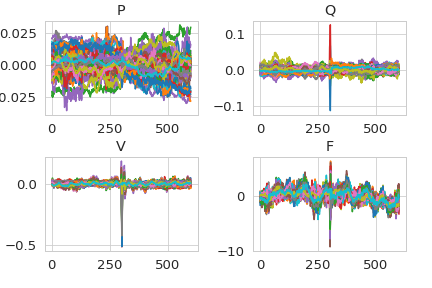}
    \end{subfigure}\hfill
    \begin{subfigure}{0.33\linewidth}
        \includegraphics[width=\linewidth]{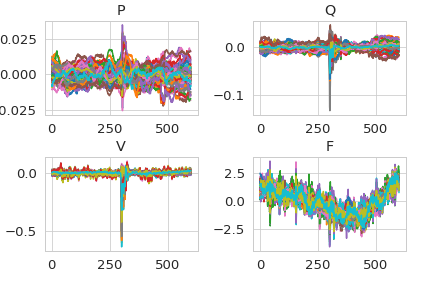}
    \end{subfigure}\hfill
    \begin{subfigure}{0.33\linewidth}
        \includegraphics[width=\linewidth]{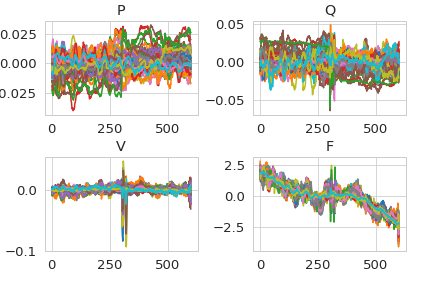}
    \end{subfigure}\hfill
    
    \caption{Samples of generated full event tensors designated as ``voltage events" caused by lightning strikes.
    }
    \label{fig:vgensamples_lightning}
\end{figure*}

\begin{figure*}[h]
    \centering
    
    \begin{subfigure}{0.33\linewidth}
        \includegraphics[width=\linewidth]{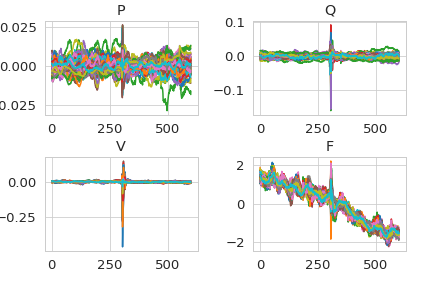}
    \end{subfigure}\hfill
    \begin{subfigure}{0.33\linewidth}
        \includegraphics[width=\linewidth]{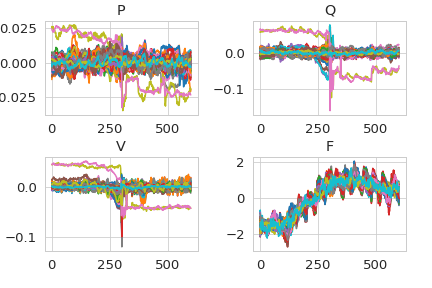}
    \end{subfigure}\hfill
    \begin{subfigure}{0.33\linewidth}
        \includegraphics[width=\linewidth]{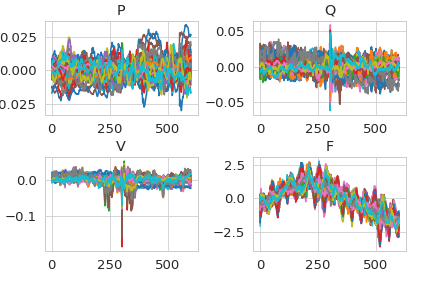}
    \end{subfigure}\hfill
    
    \caption{Samples of generated full event tensors designated as ``voltage events" caused by line trips.
    }
    \label{fig:vgensamples_trips}
\end{figure*}

\begin{figure*}[h]
    \centering
    
    \begin{subfigure}{0.33\linewidth}
        \includegraphics[width=\linewidth]{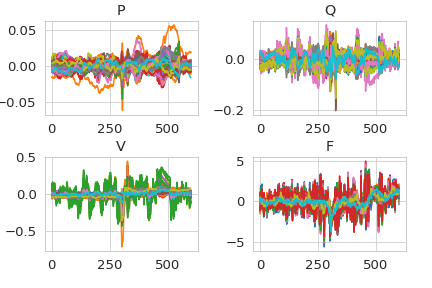}
    \end{subfigure}\hfill
    \begin{subfigure}{0.33\linewidth}
        \includegraphics[width=\linewidth]{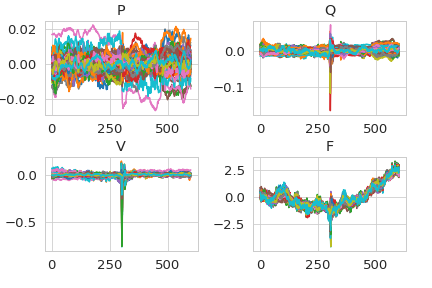}
    \end{subfigure}\hfill
    \begin{subfigure}{0.33\linewidth}
        \includegraphics[width=\linewidth]{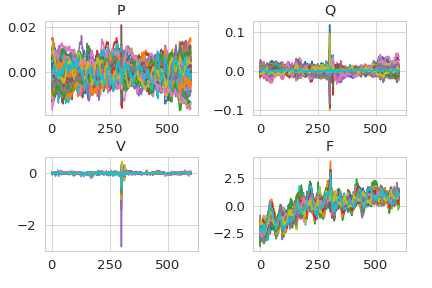}
    \end{subfigure}\hfill
    
    \caption{Samples of generated full event tensors designated as ``voltage events" caused by wind. 
    }
    \label{fig:vgensamples_wind}
\end{figure*}

\begin{figure*}[h]
    \centering
    
    \begin{subfigure}{0.33\linewidth}
        \includegraphics[width=\linewidth]{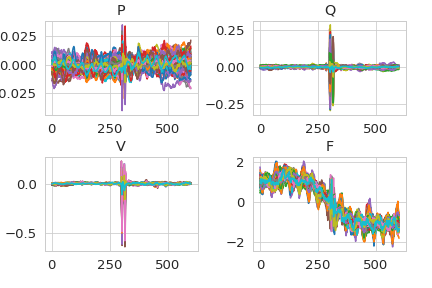}
    \end{subfigure}\hfill
    \begin{subfigure}{0.33\linewidth}
        \includegraphics[width=\linewidth]{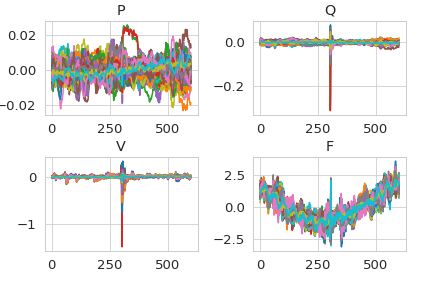}
    \end{subfigure}\hfill
    \begin{subfigure}{0.33\linewidth}
        \includegraphics[width=\linewidth]{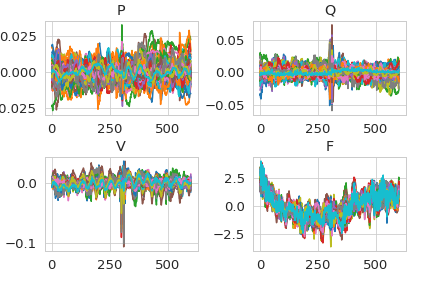}
    \end{subfigure}\hfill
    
    \caption{Samples of generated full event tensors designated as ``voltage events" caused by equipment failures. 
    }
    \label{fig:vgensamples_equipment}
\end{figure*}

Sample images of voltage events in pmuBAGE are demonstrated in Figures \ref{fig:vgensamples_lightning} through \ref{fig:vgensamples_equipment}. Here, the granularity of the event labels is increased into further groups of ``lightning strikes", ``line trips", ``wind", and ``equipment failures".

\begin{figure*}[h]
    \centering
    
    \begin{subfigure}{0.33\linewidth}
        \includegraphics[width=\linewidth]{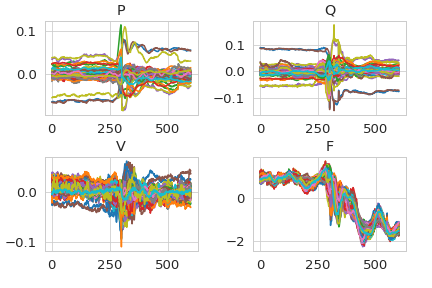}
    \end{subfigure}\hfill
    \begin{subfigure}{0.33\linewidth}
        \includegraphics[width=\linewidth]{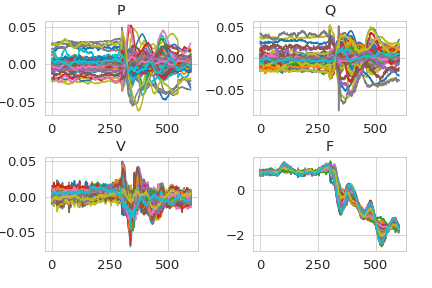}
    \end{subfigure}\hfill
    \begin{subfigure}{0.33\linewidth}
        \includegraphics[width=\linewidth]{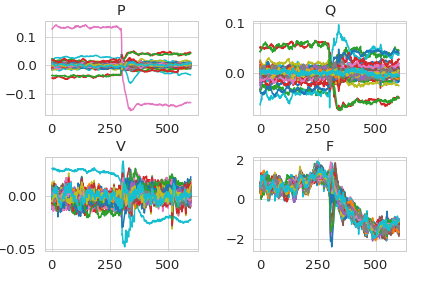}
    \end{subfigure}\hfill
    
    \begin{subfigure}{0.33\linewidth}
        \includegraphics[width=\linewidth]{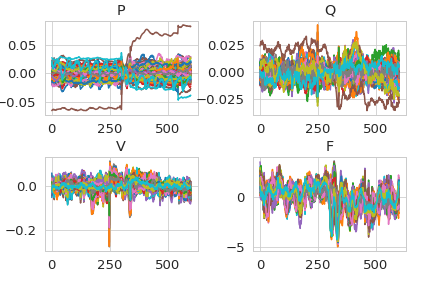}
    \end{subfigure}\hfill
    \begin{subfigure}{0.33\linewidth}
        \includegraphics[width=\linewidth]{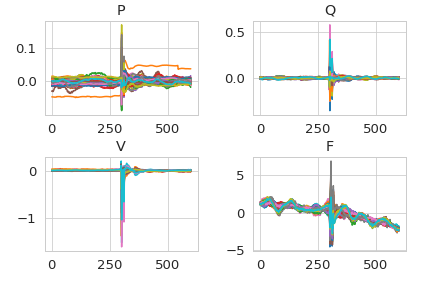}
    \end{subfigure}\hfill
    \begin{subfigure}{0.33\linewidth}
        \includegraphics[width=\linewidth]{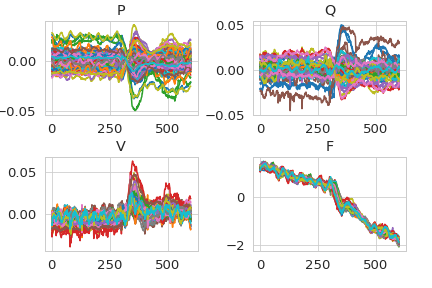}
    \end{subfigure}\hfill
    
    \caption{Samples of generated full event tensors designated as ``frequency events" caused by generator trips.
    }
    \label{fig:fgensamples_trip}
\end{figure*}

\begin{figure*}[h]
    \centering
    
    \begin{subfigure}{0.33\linewidth}
        \includegraphics[width=\linewidth]{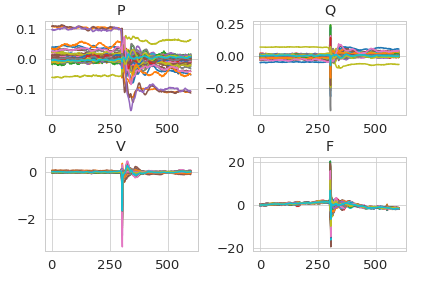}
    \end{subfigure}\hfill
    \begin{subfigure}{0.33\linewidth}
        \includegraphics[width=\linewidth]{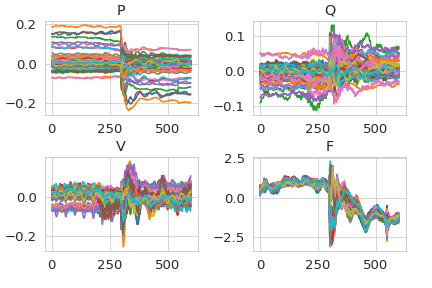}
    \end{subfigure}\hfill
    \begin{subfigure}{0.33\linewidth}
        \includegraphics[width=\linewidth]{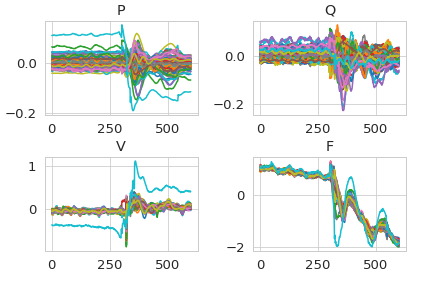}
    \end{subfigure}\hfill
    
    \begin{subfigure}{0.33\linewidth}
        \includegraphics[width=\linewidth]{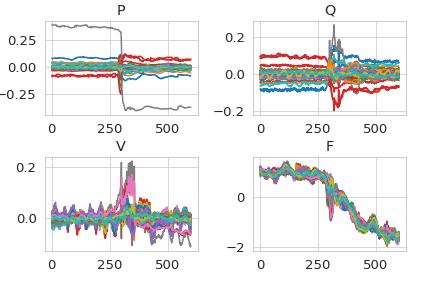}
    \end{subfigure}\hfill
    \begin{subfigure}{0.33\linewidth}
        \includegraphics[width=\linewidth]{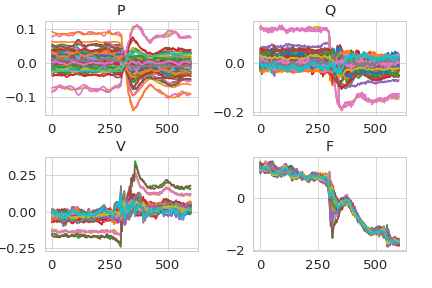}
    \end{subfigure}\hfill
    \begin{subfigure}{0.33\linewidth}
        \includegraphics[width=\linewidth]{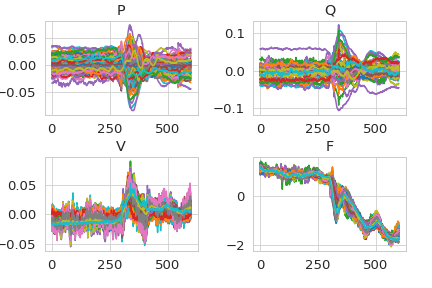}
    \end{subfigure}\hfill
    
    \caption{Samples of generated full event tensors designated as ``frequency events" caused by on-premises generator equipment failures.
    }
    \label{fig:fgensamples_equipment}
\end{figure*}

Sample images of Frequency Events in pmuBAGE are demonstrated in Figures \ref{fig:fgensamples_trip} and \ref{fig:fgensamples_equipment}. Here, the granularity of the event labels is increase into further groups of ``generator trips", and ``on premises generator equipment failures".




\end{document}